\documentclass[12pt]{article}
\begin{document}
\global\parskip 6pt
\newcommand{\be}{\begin{equation}}
\newcommand{\ee}{\end{equation}}
\newcommand{\bea}{\begin{eqnarray}}
\newcommand{\eea}{\end{eqnarray}}
\newcommand{\non}{\nonumber}

\begin{titlepage}
\hfill{hep-th/0311147} \vspace*{1cm}
\begin{center}
{\Large\bf Brane-worlds in T-dual Bulks}\\
\vspace*{2cm} Massimiliano Rinaldi\footnote{E-mail: massimiliano.rinaldi@ucd.ie}\\
\vspace*{.5cm}
{\em Department of Mathematical Physics\\
University College Dublin\\
Belfield, Dublin 4, Ireland}\\
\vspace{2cm}

\begin{abstract}

\noindent We consider brane-world models with a Schwarzschild-AdS
black hole bulk. In the particular case of a flat black hole
horizon geometry, we study the behaviour of the brane cosmological
equations when T-duality transformations act on the bulk. We find
that the scale factor is inverted and that either the Friedmann
equation or the energy conservation equation are unchanged.
However, these become both invariant if we include a tension in
the brane action. In this case, the T-duality in the bulk is
completely equivalent to the scale factor duality on the brane.

\end{abstract}

\vspace{1cm} November 2003
\end{center}
\end{titlepage}
\vspace{1.5 cm}

\section {Introduction}
\noindent One important lesson learnt from string theory is that
even though two or more theories appear to be very different, they
might actually be the same theory seen from different points of
view. With this spirit in mind, in this paper we consider two
aspects of string cosmology which have attracted the attention of
many researchers in the last decade: Pre-Big Bang scenario and
brane-world models. The former is a theory essentially based on
the $O(d,d)$ symmetry group of certain cosmological backgrounds
which appear in low energy string theory (for a comprehensive
review see \cite{Gasperini}). One of the element of the symmetry
group manifests itself through the invariance of the equations of
motion under the inversion of the scale factor. This element
corresponds to a T-duality transformation along the time
direction, also known as scale factor duality. Together with the
time-reversal symmetry of the field equations, scale factor
duality smoothly connect a pre-big bang phase of growing curvature
to the present expanding phase of decreasing curvature. The
transition between the two phases is expected to occur in a
non-perturbative string theory regime, in such a way that the
standard cosmology is recovered after the Big Bang. This
requirement is often called the graceful exit problem, since the
details are still under investigation.

The second idea consist in considering our Universe as a
3-dimensional surface embedded in a 5-dimensional bulk space-time
(good reviews can be found in \cite{Langlois,Brax}). Matter is
confined on the brane, together with all the fundamental forces
except gravity, which is a 5-dimensional field. Nevertheless,
gravity on the brane is approximately Newtonian, and this property
holds even when the extent of the extra-dimension is infinite
\cite{RS}. Among the appealing features of this model there is
also a possible explanation for the huge hierarchy between the
electro-weak scale and the Planck scale. Moreover, if the brane is
not static, an observer living on it then observes an effective
4-dimensional cosmology described by a modified Friedmann
equation, which, under appropriate conditions, matches with the
standard one at late times.

A connection between brane-world models and Pre-Big Bang scenario
is expected for at least one reason, i.e. symmetry. Indeed, on one
hand we have the Pre-Big Bang scenario which is based on a large
$O(d,d)$ symmetry group acting on the action. On the other hand,
we have branes embedded in bulks which, in most cases, have
themselves a large number of isometries. For example, in the
Randall-Sundrum model proposed in \cite{RS}, the brane glues
together two slices of anti-de Sitter space, i.e. a maximally
symmetric space. Therefore, it could be interesting to see what
happens to the effective cosmology on the brane under the action
of the isometry group on the bulk. As a concrete example, in this
paper we examine two space-times which are related by T-duality,
and which can host a brane with a flat FLRW induced metric. As we
will shortly see, the duality transformations on the bulk induce
the inversion of the scale factor on the brane. This simple fact
leads to think of a possible connection with the scale factor
duality of Pre-Big Bang scenario. We will see that this connection
exists, provided that some conditions are satisfied by the brane
matter Lagrangian.

In the next section, we briefly introduce two space-times related
by T-duality in the context of type IIA and type IIB string
theory. Their compactification to five dimensions leads to two
black hole metrics, one of these being the well known
AdS-Schwarzschild black hole with toroidal horizon. In the third
section we review the cosmological equations induced on the brane
moving in the AdS-Schwarzschild black hole bulk. These are then
compared to the ones obtained on the brane embedded in the dual
bulk in the third section. Finally, in the forth section we
examine a simple solution to the dual Friedmann equations, and we
conclude with a summary and few remarks. We also add an appendix,
where we show how the Lanczos-Israel junction conditions in string
frame can be obtained from the ones written in Einstein frame by
means of a conformal transformation.

\section {The T-dual backgrounds}

We consider the generalization of the $AdS_5 \times S^5$
compactification of type IIB string theory, obtained by replacing
the $AdS_5$ space-time with a 5-dimensional topological black hole
(\cite{Schwarz}, see also \cite{Strominger,Duff}). The
$10$-dimensional metric is \be\label{10tormetric}
ds^2=-f(r)dt^2+f(r)^{-1}dr^2+\frac{r^2}{l^2}d\Sigma_3^2+l^2d\Omega_5^2,
\ee and the function $f(r)$ is defined by \be\label{effe}
f(r)=k+\frac{r^2}{l^2}-\frac{\mu}{r^2}, \ee where $l$ and $\mu$
are constant. The parameter $k$ determines the geometry of the
horizon and from now on we set $k=0$, i.e. we consider only flat
(or toroidal) horizons with metric
$d\Sigma_3^2=\delta_{ij}dx^idx^j$. The metric (\ref{10tormetric})
must be supplemented with an anti-self Hodge dual $5$-form
\be\label{5form} F^{(5)}=-{}^{\star}F^{(5)}, \ee in order to
satisfy the fundamental equation \cite{Schwarz} \be
R_{MN}=\frac{1}{6\cdot4^2}
F_M^{\hspace{3mm}A_2A_3A_4A_5}F_{NA_2A_3A_4A_5}. \ee The
10-dimensional space-time (\ref{10tormetric}) can be easily
compactified, yielding a 5-dimensional AdS-Schwarzschild
topological black hole with flat or toroidal horizon \cite{danny}.
The metric reads\be\label{tormetric}
ds^2=-f(r)dt^2+f(r)^{-1}dr^2+\frac{r^2}{l^2}\delta_{ij}dx^idx^j,\ee
where now $l$ is related to the 5-dimensional cosmological
constant by $\Lambda=-(6/l^2)$, and $\mu$ is the mass of the black
hole. These fields are solutions to the equations of motion
derived from the effective action \be\label{IIbaction}
{}^{(5)}S=\frac{1}{2}\int dx^5\sqrt{g}\left[R-2\Lambda\right].\ee

In \cite{Max}, it was shown that the solution (\ref{10tormetric})
and (\ref{5form}) can be mapped by T-duality transformations into
new low energy solutions of type IIA or IIB string theory. In
particular, by applying three T-duality transformations along the
horizon coordinates, one obtains a type IIA background with metric
and dilaton given by
\begin{eqnarray}\label{dual3} ds^2&=&-f(r)dt^2+f(r)^{-1}dr^2+
\frac{l^2}{r^2}\delta_{ij}dx^idx^j+l^2d\Omega_5^2,
\\\nonumber\\\label{dualdilaton}
e^{-2\phi}&=&\frac{r^6}{l^6},
\end{eqnarray} and forms
\begin{eqnarray}\label{forms3}
F^{(2)}_{\quad\mu_1\mu_2}&=&\frac{4r^3}{l^4}\varepsilon_{\mu_1\mu_2},\qquad
\mu\neq x_i, \\\nonumber\\ F^{(8)}_{\quad A_1\ldots
A_8}&=&-\frac{4r^3}{l^4}\varepsilon_{A_1\ldots A_8},\qquad A_i\neq
t,r. \end{eqnarray} In addition to these, a $B$-field can be
switched on by boosting the metric (\ref{10tormetric}) along one
of the horizon direction before applying T-duality \cite{Max}. For
simplicity, here we set the $B$-field to zero. These fields solve
the equations of motion derived from the type IIA low energy
string action \be S=\frac{1}{2}\int
d^{10}x\sqrt{-G}\left\{e^{-2\phi}\left[R+4(\nabla\phi)^2\right]-\frac{1}{4}F_{MN}F^{MN}\right\},
\ee which can be easily compactified to 5 dimensions. Indeed, if
we assume that $S^5$ has unit volume, we then find
\be\label{redaction} S=\frac{1}{2}\int
d^{5}x\sqrt{-g}\left\{e^{-2\phi}\left[R+4(\nabla\phi)^2+\frac{20}{l^2}\right]-\frac{1}{4}F_{\mu\nu}F^{\mu\nu}\right\},
\ee where \bea\label{redmetric}
g_{\mu\nu}dx^{\mu}dx^{\nu}&=&-f(r)dt^2+f(r)^{-1}dr^2+
\frac{l^2}{r^2}\delta_{ij}dx^idx^j\\\nonumber\\\label{redfields}
F_{tr}&=&\frac{4r^3}{l^4}, \qquad \phi=-3\log\frac{r}{l}.\eea

\section{The brane-world}

We now examine the behaviour of a 4-dimensional brane radially
moving in the 5-dimensional compactifications of these T-dual
backgrounds. In the case of a bulk with metric (\ref{tormetric}),
the 4-dimensional cosmological equations are well known
\cite{Langlois,Brax}. If we add to the bulk action
(\ref{IIbaction}) a boundary term of the form \be\label{branelag}
S_{brane}=\int dx^4\sqrt{h}\left(L-K\right),\ee where $L$ and $K$
are respectively the matter Lagrangian and the trace of the
extrinsic curvature, and if assume a $\textbf{Z}_2$ symmetry about
the brane, then we must impose the Lanczos-Israel junction
conditions \be\label{junction}
K_{\mu\nu}=-\frac{1}{2}\left(\tau_{\mu\nu}-\frac{1}{3}\tau
h_{\mu\nu}\right).\ee  This equation relates the brane matter
stress tensor $\tau_{\mu\nu}$ to the extrinsic curvature
$K_{\mu\nu}$ defined by \be K^{\mu\nu}=\nabla^{\mu}n^{\mu},\ee
where $n^{\mu}$ is the unit vector normal to the brane and
pointing into the bulk. The stress tensor components are
$\tau^{\mu}_{\;\nu}=diag(-\rho,p,p,p)$, where $\rho$ and $p$ are
respectively the energy density and pressure. The vector $n^{\mu}$
is normalized so that the induced metric on the brane is given by
\be
h_{\mu\nu}dx^{\mu}x^{\nu}=-d\tau^2+a(\tau)^2\delta_{ij}dx^idx^j,\ee
where $a(\tau)=r(\tau)/l$ is the scale factor and $\tau$ is an
affine parameter, usually identified with the cosmic time measured
on the brane. With these settings, we can combine the $(\tau\tau)$
and the $(ij)$ components of Eq.~(\ref{junction}) and obtain
\cite{Langlois,Brax}
\begin{eqnarray}\label{friedmann} \frac{\dot{r}^2}{r^2} &=&
\frac{\rho^2}{36}-\frac{1}{l^2}+\frac{\mu}{r^4}\\\nonumber\\\label{cons}
\dot{\rho} &=& -3\frac{\dot r}{r}\rho(\omega+1), \end{eqnarray}
where the dot stands for differentiation with respect $\tau$, and
where we assume that $p=\omega\rho$. These expressions can be
interpreted respectively as the Friedmann and the energy
conservation equations by a 4-dimensional observer living on the
brane, provided that we define the Hubble function \be\label{acca}
H=\frac{\dot a}{a}=\frac{\dot{r}}{r}.\ee We see that the Hubble
function in Eq.~(\ref{friedmann}) is proportional to $\rho$,
instead of $\sqrt{\rho}$, as in standard cosmology. However
\cite{Langlois,Brax}, if we add a tension $V$ to the Lagrangian of
the brane action (\ref{branelag}), then $\rho\rightarrow \rho+V$,
and the Friedmann equation becomes \be\label{friedmann2}
H^2=\frac{1}{18}V\rho+\frac{\rho^2}{36}+\frac{V^2}{36}
+\frac{\mu}{r^4}-\frac{1}{l^2}.\ee Therefore, when $V\gg \rho$, we
can neglect the term proportional to $\rho^2$ and recover the
standard cosmology up to the so-called dark radiation term
$\mu/r^4$, and the constant \be C=\frac{V^2}{36}-\frac{1}{l^2}.\ee
If we set $\omega=-1$ then, according to Eq.~(\ref{cons}), $\rho$
is constant. If we also set $C=0$ (fine-tuning), then $\mu$ must
necessarily vanish \cite{max2}, and we recover the standard
Randall-Sundrum scenario with one (static) brane \cite{RS}.

The metrics (\ref{tormetric}) and (\ref{redmetric}) are related by
the inversion of the $(ij)$ components. It is then clear that if
we embed a 4-dimensional brane in the dual bulk space-time
(\ref{redmetric}), then the induced metric will be
\be\label{indmetric} ds^2=-d\tau^2+\tilde
{a}(\tau)^2\delta_{ij}dx^idx^j,\ee where now
\be\label{dualscalefactor}
\tilde{a}(\tau)=\frac{l}{r(\tau)}=\frac{1}{a(\tau)}\ee is the new
scale factor. Therefore, the T-duality transformations on the bulk
generate the inversion of the scale factor of the induced metric
on the brane. The natural question is then how the effective
4-dimensional cosmology is affected. In particular, if the
cosmological equations are unchanged, then the T-duality symmetry
of the bulk corresponds to a scale factor duality symmetry on the
brane. In the next section we will see that this can indeed be the
case.

\section{The dual brane-world}

The dual fields (\ref{redmetric}) and (\ref{redfields}) are
solutions of the equations of motion derived from the action
(\ref{redaction}) written with respect to the string frame.
Therefore, in order to study the brane cosmological equations, we
need the junction conditions in string frame which read (see the
Appendix) \bea\label{junctionstring}
K_{\alpha\beta}&=&-\frac{1}{2}\Omega^{(q+1)}\left(\tau_{\alpha\beta}-\frac{\tau}{3}h_{\alpha\beta}\right)
-h_{\alpha\beta}\Omega^{-1}n^{\mu}\partial_{\mu}\Omega,\\\nonumber\\\label{junctiondilaton}
n^{\mu}\partial_{\mu}&=&-\frac{1}{2}\Omega^{(q+1)}\frac{\partial\xi}{\partial
\phi}\tau,\eea where $\Omega^2$ is the conformal factor which
relates string and Einstein frame, and $q$ determines the coupling
of the brane Lagrangian to the induced metric. In our case, the
action (\ref{redaction}) can be transformed in Einstein frame
through the conformal rescaling \be
g_{\mu\nu}=e^{4\phi/3}\tilde{g}_{\mu\nu}=\Omega^{-2}\tilde{g}_{\mu\nu},
\ee hence, according to our notations, \be \Omega=e^{-2\phi/3}.\ee
To determine the unit normal vector $n^{\mu}$, we first impose
that the brane moves along a radial geodesics with velocity
$u^{\mu}=(\dot t, \dot r,0,0,0)$, where the dot stands for
differentiation with respect to the affine parameter $\tau$
\cite{Langlois,Brax}. The bulk metric (\ref{redmetric}) can then
be written as \be
ds^2=-\left[f(r)\dot{t}^2-f(r)^{-1}\dot{r}^2\right]d\tau^2+
\tilde{a}^2(\tau)\delta_{ij}dx^idx^j, \ee and, if we impose the
normalization $f(r)\dot{t}^2-f(r)^{-1}\dot{r}^2=1$, we then obtain
the induced metric (\ref{indmetric}). Finally, the condition
$n^{\mu}u_{\mu}=0$ leads to $n_{\alpha}=(\dot{r},-\dot{t},0,0,0)$.
With these settings, the $(ij)$ and $(\tau\tau)$ components of
Eq.~(\ref{junctionstring}) read respectively \begin{eqnarray}
\frac{1}{\dot{r}}\frac{d}{d\tau}\sqrt{f+\dot{r}^2}+\frac{4}{r}\sqrt{f+\dot{r}^2}&=&-\frac{1}{2}\omega\rho
e^{-\frac{2}{3}(q+1)\phi},
\\\nonumber\\\frac{1}{r}\sqrt{f+\dot{r}^2}&=&\frac{\rho}{6}e^{-\frac{2}{3}(q+1)\phi},\end{eqnarray}
where we also assumed that $\tau^{\mu}_{\;\nu}=\rho\;diag\;
(-1,\omega,\omega,\omega)$. By squaring the $(\tau\tau)$
components and using the definition (\ref{effe}) with $k=0$, we
find \be\label{dualfried}
\frac{\dot{r}^2}{r^2}=\frac{\rho^2}{36}e^{-\frac{4}{3}(q+1)\phi}-\frac{1}{l^2}+\frac{\mu}{r^4}.\ee
Then, with the explicit form of the dilaton (\ref{redfields}) and
Eq.~(\ref{dualfried}), the $(\tau\tau)$ component of
Eqs.~(\ref{junctionstring}) can be written as \be\label{dualcons}
\dot{\rho}=-\rho\frac{\dot{r}}{r}\left(3\omega+2q+7\right).\ee
Finally, the junction condition (\ref{junctiondilaton}) yields
\be\label{csi} \frac{\partial \xi}{\partial
\phi}=\frac{1}{1-3\omega}.\ee These conditions are valid if we
suppose that the bulk 2-form does not couple to the matter on the
brane. Consequently, there are no junction conditions to be
imposed on the bulk form field \cite{quevedo}. According to the
definitions (\ref{acca}) and (\ref{dualscalefactor}), the Hubble
function is \be \tilde H=\frac{\dot{\tilde{a}}}{\tilde
a}=-\frac{\dot{a}}{a}=-H=-\frac{\dot{r}}{r}.\ee Then,
Eqs.~(\ref{dualfried}) and (\ref{dualcons}) reads respectively
\bea\label{dualfried2}\tilde{H}^2&=&\frac{\rho^2}{36}e^{-\frac{4}{3}(q+1)\phi}
-\frac{1}{l^2}+\frac{\mu}{l^4}\tilde{a}^4
,\\\nonumber\\\label{dualcons2}\dot{\rho}&=&\rho\tilde{H}\left(3\omega+2q+7\right)
.\eea

We see that, for $q=-1$, the Friedmann equation (\ref{dualfried})
is the same as (\ref{friedmann}), while, for $q=-2$, the
conservation equation (\ref{dualcons}) is equal to
Eq.~(\ref{cons}). Hence, under T-duality transformation on the
bulk, either the Friedmann equation or the energy conservation
equation are unchanged on the brane. In particular,
Eq.~(\ref{dualcons}) describes the non-conservation of the energy
on the brane from a point of view of an observer in the bulk. To
measure the eventual energy flow from the point of view of an
observer living on the brane, we must write this equation in the
conformal frame, usually called the Jordan frame, related to the
metric $\gamma_{\mu\nu}$ (see the Appendix and \cite{Langlois}).
Usually, through the conformal transformation \be
\gamma_{\mu\nu}=e^{2\xi(\phi)}h_{\mu\nu}, \ee we ``distort'' the
brane in the bulk in such a way that the dilaton flux is
tangential to brane itself, hence its contribution to the energy
density vanishes\footnote{Thanks to P. Watts for suggesting this
interpretation.}. In the case of branes embedded in asymptotically
$AdS$ bulks with dilaton field, one always find that, in the
Jordan frame, the energy is conserved \cite{Langlois,Brax}.
However, this does not happen in our model. Indeed, if we follow
\cite{Langlois} and we replace \be \rho\rightarrow
e^{-4\xi(\phi)}\rho, \qquad \tilde{a}\rightarrow
e^{\xi(\phi)}\tilde{a}, \qquad dt\rightarrow e^{\xi(\phi)}dt, \ee
in Eq.~(\ref{dualcons2}), we obtain \be
\dot\rho+\rho\tilde{H}\left[\frac{3\xi'}{1+3\xi'}(4+\beta)-\beta\right]=0,
\ee where $\xi'=\partial_{\phi}\xi(\phi)$ and
$\beta=3\omega+2q+7$. By using Eq.~(\ref{csi}), we find that \be
\frac{3\xi'}{1+3\xi'}(4+\beta)-\beta=3(\omega+1) \quad
\Leftrightarrow \quad q=-\frac{1}{2}(7+6\omega). \ee This means
that the energy on the brane, measured with respect to the Jordan
frame, is \emph{not} in general conserved. The reason for this
``anomaly'' becomes clear once we remember that the junction
conditions depend on the induced metric $h_{\alpha\beta}$ which is
related to the bulk metric. The latter is a solution to the bulk
equations of motion which also include the $2$-form. Hence, even
if we suppose that it does not couple to the matter, the 2-form
does affect the brane dynamics through the bulk equations of
motion. In other words, the 2-form flux through the brane does not
in general vanish, even in the Jordan frame.

This suggests that, if we chose a more appropriate brane
Lagrangian, the energy might be conserved in the Jordan frame or,
more importantly, even in the conformal frame defined by
$h_{\mu\nu}$. In this case, both cosmological equations would be
invariant under T-duality in the bulk. Therefore, we generalise
the brane Lagrangian by considering the total energy density on
the brane as sum the of the brane matter energy density and a
\emph{time dependent} tension, i.e. we set
$\rho=\tilde{\rho}+V(\tau)$. It is then easy to show that
Eq.~(\ref{dualcons}) takes the form \be
\dot{\tilde{\rho}}=-3\tilde{\rho}\frac{\dot{r}}{r}(\omega+1)=\tilde{\rho}
\tilde{H}(\omega+1),\ee provided that we use the relation
$\dot{\phi}=3\tilde{H}$ and we assume that \be V(\tau)=e^{a\phi},
\ee where $a$ is an arbitrary constant. With this form, the
time-dependent tension can also be interpreted as an effective
dilatonic potential on the brane (similar solutions were found in
\cite{cai}, see also \cite{quevedo2}). Also, note that the
Friedmann equation (\ref{dualfried}) assume the same form of
Eq.~(\ref{friedmann2}). Therefore we conclude that, when $q=-1$
and when there is a dilatonic potential on the brane, \emph{both}
brane cosmological equations are invariant under T-duality in the
bulk.

\section{Duality of brane dynamics}

According to the results of the previous section, to each
expanding solution to the Friedmann equation on the brane, there
exists a dual contracting one, in strict analogy with the Pre-Big
Bang scenario. As an example, let us consider the solutions to
Eqs.~(\ref{friedmann}) and (\ref{cons}) for $\mu=0$. In this case,
Eq.~(\ref{cons}) can be easily solved, yielding \be
\rho=\rho_0a^{-3(\omega+1)},\ee $\rho_0$ being an arbitrary
constant. If we assume that the total energy density on the brane
is given by $\rho_{tot}=V+\rho$ and impose the fine-tuning
$6/l=V$, then the solution to Eq.~(\ref{friedmann2}) reads
 \be
\tilde{a}^{-3(w+1)}=a^{3(\omega+1)}=At+Bt^2,\ee where $A$ and $B$
are positive constants \cite{Langlois}. These solutions are valid
also when $\mu\neq 0$, provided that $t$ is large. Indeed, in this
case the scale factor $a=r/l$ of the type IIB background is large
and the mass term in Eq.~(\ref{friedmann2}) is negligible.  This
situation corresponds to a brane moving in the asymptotic region
of the black hole, where the space-time is locally anti-de Sitter
\cite{max2}. In the type IIA background, for large values of $t$,
the scale factor becomes very small, and in Eq.~(\ref{dualfried2})
the mass term tends again to zero. Since $\phi=3\log \tilde{a}$,
we see that, for large $t$, we have large values of the effective
string coupling $e^{-2\phi}$. Hence, the brane moving in the dual
background of type IIA enters the strong coupling regime at late
times (i.e. large $r$). In the type IIB background there is no
dilaton, however a breakdown of the model occurs when the brane
approaches the horizon and the tension diverges \cite{max2}.
Therefore, even in this simplified model we can see the duality
between physics at small and large distances typical of T-duality.

\section{Conclusions}

We have considered two 4-dimensional branes moving respectively in
two bulks related by T-duality. The branes have reciprocal scale
factors and we found that either the effective Friedmann equation
or the energy conservation equation are unchanged, according to
which form of the coupling between brane matter and induce metric
we choose. However, we showed that if we add to the brane
lagrangian a dilatonic potential, then it is possible to have both
equation invariant under duality. Therefore, we have an exact
equivalence between scale factor duality on the brane and
T-duality in the bulk. To obtain these result we have used the
junction conditions in string frame obtained, by means of a
conformal transformation, from the junction conditions in Einstein
frame. Finally, we analyzed a simple solution to the Friedmann
equation, and we showed how the typical correspondence of
T-duality between large and small distances emerges in the brane
dynamics.

These results may have interesting applications in the context of
the Pre-Big Bang scenario, in particular in relation with the
graceful exit problem. Also, the equivalence between bulk and
brane duality might be extended by means of non-Abelian
T-dualities \cite{moha1,moha2} or to more general string
backgrounds. We think that all these aspects deserve further
investigations. \vspace{0.5cm}

\noindent {\bf \large Acknowledgements} We wish to thank P. Watts
for valuable discussions.

\appendix

\section*{Appendix: junction conditions in string frame}

The Lanczos-Israel matching conditions in string frame have been
found in various works (see for example \cite{Visser,Foffa}), by
extremizing the bulk action implemented by the appropriate
boundary terms. Here, we propose a much simpler approach based on
the conformal mapping between the action written in Einstein frame
and in string frame. Consider the generic action in Einstein frame
with a boundary term \be\label{eframeaction}
\tilde{S}=\frac{1}{2}\int
dx^5\sqrt{\tilde{g}}\left\{\tilde{R}-\frac{4}{3}(\partial
\phi)^2+\tilde{L}_{bulk}\right\}+\int
dx^4\sqrt{\tilde{h}}\tilde{L}_{brane},\ee where $\tilde{L}_{bulk}$
and $\tilde{L}_{brane}$ denote generic matter Lagrangians in the
bulk and in the brane respectively, $\phi$ is the dilaton field,
$\tilde{h}_{\mu\nu}=\tilde{g}_{\mu\nu}-\tilde{n}_{\mu}\tilde{n}_{\nu}$
is the induced metric on the brane, and $\tilde{n}^{\mu}$ is the
unit vector normal to the brane. By assuming \textbf{Z}$_2$
symmetry about the brane, the junction conditions to be satisfied
by the metric read \cite{Lanczos1}-\cite{Israel}
\be\label{ejunctionmetric}
\tilde{K}_{\mu\nu}=-\frac{1}{2}\left(\tilde{\tau}_{\mu\nu}-\frac{1}{3}\tilde{\tau}\tilde{h}_{\mu\nu}\right),
\ee where $\tilde{\tau}_{\mu\nu}$ is the stress tensor of the
brane matter, and
$\tilde{K}^{\mu\nu}=\tilde{\nabla}^{\mu}\tilde{n}^{\nu}$ is the
extrinsic curvature of the brane. We also assume that the matter
fields confined on the brane are coupled to a metric
$\tilde{\gamma}_{\mu\nu}$, conformally related to the induced
metric through \be
\tilde{\gamma}_{\alpha\beta}=e^{2\tilde{\xi}(\phi)}\tilde{h}_{\alpha\beta}.\ee
Then, the dilaton field must satisfy the junction condition
\cite{Langlois} \be\label{dilatonjunct}
\tilde{n}^{\mu}\tilde{\nabla}_{\mu}\phi=-\frac{1}{2}\frac{\partial\tilde{\xi}}{\partial
\phi}\tilde{\tau}.\ee The action can be written in string frame by
means of a conformal rescaling of the metric
\be\label{confrescaling} \tilde{g}_{\mu\nu}=\Omega^2g_{\mu\nu},\ee
where $\Omega$ is in general a function of the dilaton field. If
we rescale the induced metric and the unit normal vector according
to \be \tilde{h}_{\mu\nu}=\Omega^2h_{\mu\nu},\qquad
\tilde{n}_{\alpha}=\Omega n_{\alpha},\ee then we ensure that
$n_{\mu}n^{\mu}=1$ and that $h_{\alpha}^{\;\;\;\beta}$ is a
projection operator \cite{Hawk}. Note also that
$\tilde{h}_{\alpha}^{\;\;\;\beta}=h_{\alpha}^{\;\;\;\beta}$, and
$h_{\mu\nu}=g_{\mu\nu}-n_{\mu}n_{\nu}$. By using the
transformation law for the connection coefficients \cite{Hawk} \be
\tilde{\Gamma}_{\;\;\mu\nu}^{\alpha}=\Gamma_{\;\;\mu\nu}^{\alpha}+\Omega^{-1}
\left(\delta^{\alpha}_{\;\mu}\partial_{\nu}\Omega+\delta^{\alpha}_{\;\nu}\partial_{\mu}\Omega-
g_{\mu\nu}g^{\alpha\beta}\partial_{\beta}\Omega\right),\ee we find
that the extrinsic curvature in string frame reads \be
\tilde{K}_{\alpha\beta}=\Omega
K_{\alpha\beta}+h_{\alpha\beta}n^{\mu}\partial_{\mu} \Omega.\ee In
order to write the right-hand side of Eq.~(\ref{ejunctionmetric})
in string frame, we need to know how the Lagrangian
$\tilde{L}_{brane}$ transforms under the conformal rescaling
(\ref{confrescaling}). The transformation will in general depend
on the field content of the brane Lagrangian, and here we simply
suppose that, under the transformation (\ref{confrescaling}), \be
\tilde{L}_{brane}=\Omega^q\;L_{brane},\ee for some real number
$q$. Consequently, if we define in Einstein frame
\be\label{einsteinstress}
\tilde{\tau}^{\mu\nu}=\frac{2}{\sqrt{\tilde{h}}}\frac{\delta
(\sqrt{\tilde{h}}\;\tilde{L})}{\delta \tilde{h}_{\mu\nu}},\ee then
the conformal transformation of the brane stress tensor reads \be
\tilde{\tau}^{\mu\nu}=\Omega^{(q-2)}\;\tau^{\mu\nu}.\ee
Alternatively \cite{Langlois}, we can define the brane stress
tensor in Einstein frame according to the conformal metric
$\tilde{\gamma}_{\alpha\beta}$ (often called the Jordan frame) to
which brane matter fields couple to, i.e.
\be\label{jordaneinsteinstress}
{}^{(\gamma)}\tilde{\tau}^{\mu\nu}=\frac{2}{\sqrt{\tilde{\gamma}}}\frac{\delta
(\sqrt{\tilde{h}}\;\tilde{L})}{\delta \tilde{\gamma}_{\mu\nu}}.\ee
Then, according to Eq.~(\ref{confrescaling}), in Einstein frame
the two stress tensors (\ref{einsteinstress}) and
(\ref{jordaneinsteinstress}) are related by \be
{}^{(\gamma)}\tilde{\tau}^{\mu}_{\;\;\nu}=e^{-4\tilde{\xi}(\phi)}\tilde{\tau}^{\mu}_{\;\;\nu}.\ee
In string frame, we have the analogous equivalence \be
{}^{(\gamma)}\tau^{\mu}_{\;\;\nu}=e^{-4\xi(\phi)}\tau^{\mu}_{\;\;\nu},\ee
provided that we impose
\be\gamma_{\alpha\beta}=e^{2\xi(\phi)}h_{\alpha\beta}.\ee Hence,
consistency requires that \be
\tilde{\gamma}_{\alpha\beta}=e^{2\tilde{\xi}(\phi)}\Omega^2\gamma_{\alpha\beta}e^{-2\xi(\phi)}.\ee
If we make the choice $\tilde{\xi}(\phi)=\xi(\phi)$, then
$\tilde{\gamma}_{\mu\nu}$ and $\tilde{h}_{\mu\nu}$ transform in
the same way under the rescaling (\ref{confrescaling}), i.e. \be
\tilde{\gamma}_{\mu\nu}=\Omega^2\;\gamma_{\mu\nu},\quad
\Longleftrightarrow \quad
\tilde{h}_{\mu\nu}=\Omega^2\;h_{\mu\nu}.\ee We these settings, it
is then easy to show that Eqs.~(\ref{ejunctionmetric}) and
(\ref{dilatonjunct}) in string frame read \bea
K_{\alpha\beta}&=&-\frac{1}{2}\Omega^{(q+1)}\left(\tau_{\alpha\beta}-\frac{\tau}{3}h_{\alpha\beta}\right)
-h_{\alpha\beta}\Omega^{-1}n^{\mu}\partial_{\mu}\Omega,\\\nonumber\\
n^{\mu}\partial_{\mu}&=&-\frac{1}{2}\Omega^{(q+1)}\frac{\partial\xi}{\partial
\phi}\tau.\eea If we set $q=-4$ and we define the extrinsic
curvature as $K^{\mu\nu}=-\nabla^{\mu}n^{\nu}$, we then recover
the junction conditions found in \cite{Visser}. Alternatively, for
$q=-1$ and $L$ replaced with $-L/2$, our results match with the
ones obtained in \cite{Foffa}.

\end{document}